# Automated Workflow for the Detection of Vugs

M.Q. Nasim[1], T. Maiti[2], N. Mosavat[3], P. V. Grech[4], T. Singh[2], P. Nath Singha Roy[1]

[1] Indian Institute of Technology; [2] Deepkapha; [3] Muscat University; [4] CC Energy Development

## Summary

Image logs are crucial in capturing high-quality geological information about subsurface formations. Among the various geological features that can be gleaned from Formation Micro Imager log, vugs are essential for reservoir evaluation. This paper introduces an automated Vug Detection Model, leveraging advanced computer vision techniques to streamline the vug identification process. Manual and semi-automated methods are limited by individual bias, labour-intensity and inflexibility in parameter fine-tuning. Our methodology also introduces statistical analysis on vug characteristics. Pre-processing steps, including logical file extraction and normalization, ensured standardized and usable data. The six-step vug identification methodology encompasses top-k mode extraction, adaptive thresholding, contour identification, aggregation, advanced filtering, and optional filtering for low vuggy regions. The model's adaptability is evidenced by its ability to identify vugs missed by manual picking undertaken by experts. Results demonstrate the model's accuracy through validation against expert picks. Detailed metrics, such as count, mean, and standard deviation of vug areas within zones, were introduced, showcasing the model's capabilities compared to manual picking. The vug area distribution plot enhances understanding of vug types in the reservoir. This research focuses on the identification and characterization of vugs that in turn aids in the better understanding of reservoirs.

**Automated Workflow for the Detection of Vugs**

**Introduction**

FMI images are invaluable repositories of high-quality geological metadata. Vug identification is a critical piece of information, generally in carbonate rocks, necessary for reservoir evaluation [X. Li et al., 2017]. Vugs are distinctive geological features, represented as roughly elliptical or speckled solution pores on a small scale [P. Li et al., 2015]. On a larger scale, these cavities manifest as irregular clusters, or slices, each characterized by a succinct extension [Yang et al., 2011]. Traditional semi-automated vugs detection methods, such as path opening [X. N. Li et al., 2019], are hindered by their limited flexibility, resulting in challenges associated with parameter fine-tuning. This project addresses these challenges by introducing an automated Vug Detection Model, harnessing advanced computer vision techniques to streamline and optimize the vug identification process.

Our proposed methodology automates the identification of vugs and introduces a novel dimension through statistical analysis of individual vugs. The method includes the generation of distribution plots illustrating the vug area within the zone of interest. Furthermore, our model provides detailed data on the individual area of each vug, delivering valuable insights to geologists and reservoir engineers for a more nuanced understanding of reservoir dynamics and dual tank STOOIP estimations. The goal is to enhance the efficiency and accuracy of vug detection and characterisation through FMI images, ultimately contributing to a more comprehensive understanding of subsurface geological formations.

**Dataset and Methodology**

This study utilizes data from two vertical Carbonate wells located in Oman. The dataset employed is confidential and generously provided by CC Energy Development, based in Muscat, Oman. In the pre-processing phase, logical files were extracted from the DLIS file, focusing on static FMI images for analysis. The DLIS file format, known as the "Digital Log Information Standard," serves as a standardized format for storing and exchanging well log data within the oil and gas industry. The FMI image, depth, and corresponding well-radius was extracted from the datasets. The scale of radius and depth were standardized to meters, and the entire dataset was normalized to a scale between 0 and 255. For analysis, 1-meter patches were generated, which balance interactions of local and global features. These pre-processing steps are necessary to ensure that the data is in a standardized and usable format for subsequent analysis.

The methodology for vug area identification and percentage distribution involves six distinct steps, each contributing to the precision of vug detection. Firstly, the extraction of top-k mode values is initiated by calculating the frequency of each element within the dataset, sorting them in descending order, and selecting the top-k values. This process isolates the most frequent elements, such as FMI images with no features, providing a more precise depiction of dominant patterns and characteristics within the dataset [Vyas et al., 2018]. Subsequently, adaptive thresholding is applied after extracting the top-k mode values [Roy et al., 2014]. This step subtracted each mode from the original 1-meter FMI patch, emphasizing variations and anomalies. Adaptive thresholding is then employed on the subtracted patch, determining the threshold locally for each pixel based on its neighbourhood. This technique enhances contrast and highlights specific features within the FMI patch, distinguishing geological features from background noise. The third step involves contour identification and contour analysis [Goldenberg et al., 2001], where contours are identified for the k-proposed threshold map to determine the shape and size of vugs. The contour area is a pivotal parameter for differentiating True Positives (TP) from False Positives (FP), effectively reducing the FP rate. Multiple parameters, including those for areas and circularity, control vug identification. Contour aggregation and duplication removal follow, with contours derived from the k-binary images assembled to analyse contour centroids in proximity and eliminate duplicates. Advanced filtering techniques were implemented in order to assess contours with its surrounding area. Contours exhibiting prominent contrast compared to their vicinity are retained as TP, and an additional filter evaluates the mean pixel difference within the shapes. Contours exceeding a predefined threshold are retained as TP, while others are discarded. The percentage of vugs is calculated over 10 cm intervals. The final step involves applying a filter to remove suspiciously low vuggy regions within a 10 cm zone, monitoring the overall vug percentage in the adjacent zones above and below. This optional filter considers vugs' typical clustering nature, removing potential inaccuracies based on the surrounding vug percentages. The choice of implementing this filter depends on specific use cases and requirements.

After identifying all vugs, advanced analyses are conducted to gain more nuanced insights. The primary focus is calculating each vug's area within the FMI image. This detailed analysis proves instrumental in understanding various aspects such as Vug Size Distribution, Vug Location, Vug Shape, Separate Vug Area, Individual Vug Resistivity Values, and Zonal Vug Analysis. By delving into these parameters, the study aims to unravel the spatial distribution, morphological characteristics, and electrical resistivity variations associated with individual vugs. This granular exploration of vug properties contributes to a comprehensive understanding of the subsurface geological formations and enhances the reservoir characterization process. Over a number of iterations, this proposed methodology has demonstrated its effectiveness by identifying previously overlooked vugs due to limitations in existing approaches or programs. The enhanced capability of this process to detect vugs, which conventional methods might have missed, underscores its significance in providing a more comprehensive and accurate assessment of subsurface geological features.

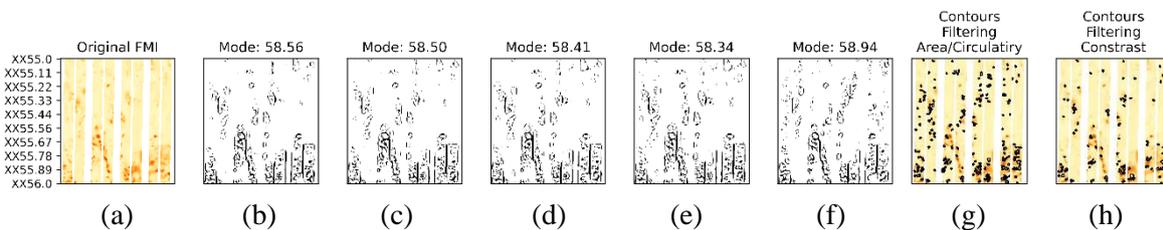

*Figure 1* *Vug identification pipeline proposed in this study starting from top-k mode extraction to advanced filtering. (a) Original 1m FMI image. (b) to (f) top-k mode extraction and subtraction which enhances the local variations in the images. (g) Area and Circularity filtering applied on the detected contours to decrease the false positives (FP). (h) Contrast based filtering applied to decrease the FP even more. To maintain the confidentiality of the data, the first two digit of the depths shown in the leftmost panel of the plot are masked.*

**Results**

Figure 1 shows the high-level pipeline of our proposed vug identification workflow. To the original FMI image (Figure 1 (a)), mode subtraction is applied to it (Figure 1 (b) to (f)) to enhance the local variation in the image. Adaptive thresholding is then applied on each of these images and generate all the possible contours. All contours are then merged from these top-k mode subtracted threshold image. Removal of duplicate contours generated while merging are also removed. And finally, several filtering techniques such as area and circularity filter are applied (Figure 1 (g)) and contrast filter (Figure 1 (h)) to decrease the false positives thereby increasing the precision of this model. Several other optional filters are also present in the methodology proposed in this study. Figure 2 presents detailed information about individual vugs in contours, providing a visual representation of their spatial distribution. To validate the accuracy of our estimated vug percentage's accuracy, results over every 10 cm were compared to expert picks, revealing consistent alignment between the two methodologies. Overall, a good correlation is observed between Predicted and GT in terms of % of vug identification. Additionally, supplementary insights were introduced, including the count of vugs and the mean and standard deviation of vug areas over each 10 cm zone. These detailed metrics were not previously available with conventional vug identification methods, showcasing the enhanced capabilities of this approach. Figure 3 is a display of one of the characteristics, vug area variability, for the same section shown in Figure 2. The distribution plot of vug areas offers geologists and reservoir engineers a holistic understanding of the types of vugs present in the reservoir, providing valuable insights for reservoir characterization and enhancing the overall reservoir evaluation process.

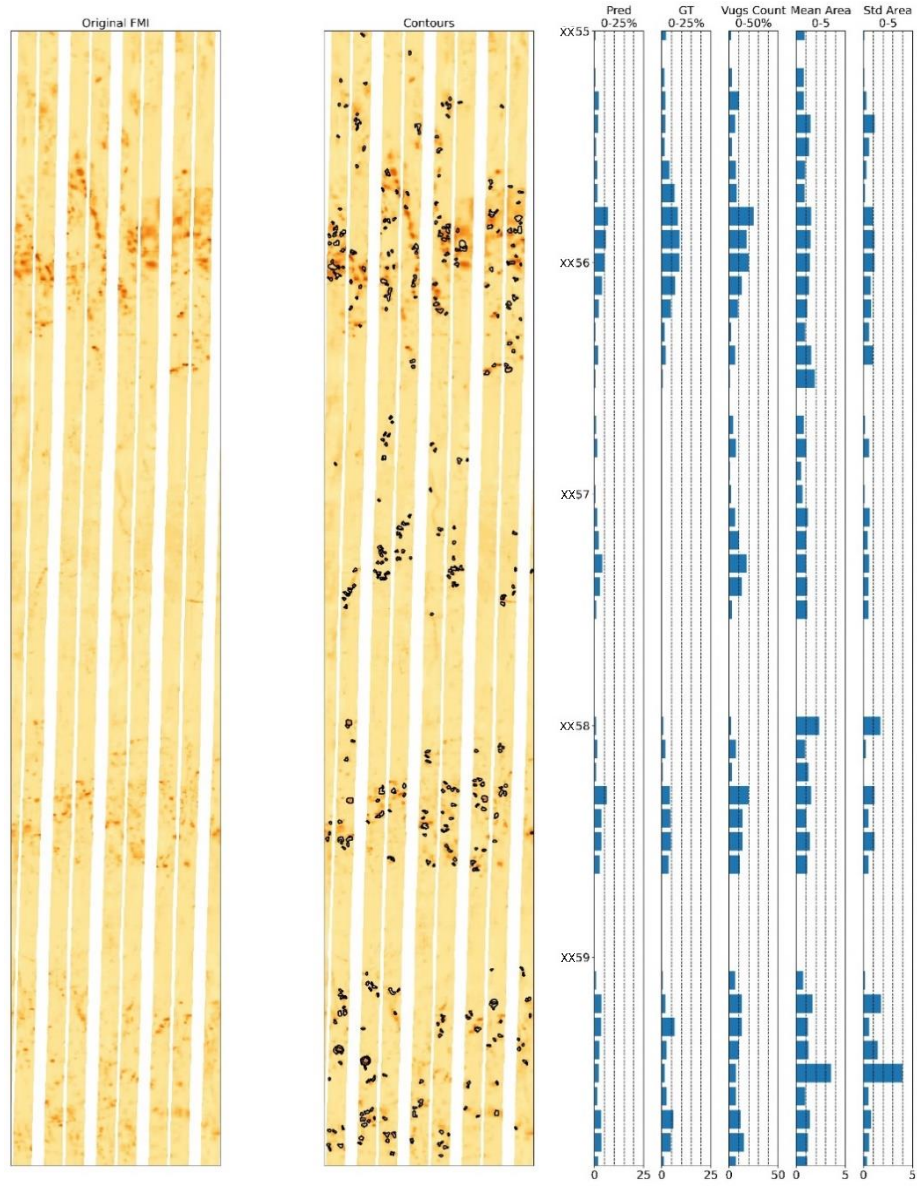

*Figure 2* Detailed contour representations reveal individual vugs, showcasing their spatial distribution, comparison with expert picks, and additional insights, including vug count, mean, and standard deviation of vug areas within each 10 cm zone, providing comprehensive metrics not available with conventional methods. To maintain the confidentiality of the data, the first two digit of the depths shown in the bar chart of the plot are masked.

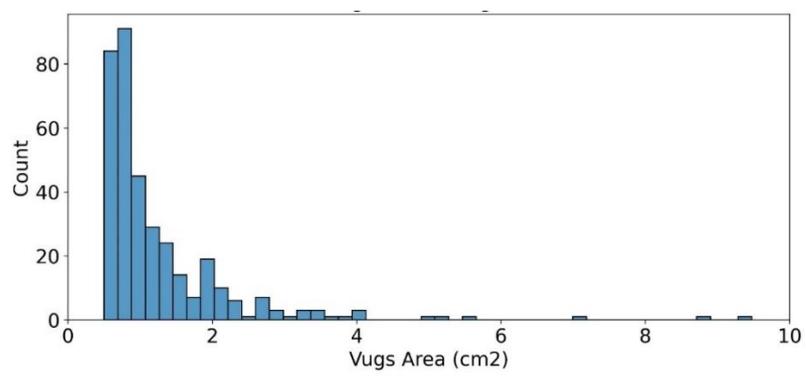

*Figure 3* Distribution plot of vug areas provides a holistic understanding, aiding geologists and reservoir engineers in characterizing the types of Vugs in the reservoir.

For a more quantitative analysis, we used mean absolute error (MAE) to assess our model's performance. An MAE of 1.21 cm$^2$ was observed compared to the expert's pick. Our focus on the interpretability of vugs identification (Figure 1), enhanced interpretation such as contours on individual vugs (Figure 2), and statistical analysis (Figure 3) can be a helpful tool that should help geologists and reservoir engineers in making quick and well-informed decisions.

**Conclusions**

Wang and Wang (2021) utilize EMD for de-noising electrical imaging logs and vug extraction. While this is effective in enhancing image clarity, it removes subtle features critical for the accurate characterization of vugs due to its inherent challenge in distinguishing between noise and fine-scale geological details. Their study does not comprehensively analyse the vug percentage and area at every 1m interval. In contrast, the method presented here can be efficiently adapted to other wells with only four parameters: min/max vug area and min/max circle ratio. Through a six-step process, from top-k mode extraction to advanced filtering, our model demonstrates precision, adaptability, and occasional overestimation, providing opportunities for refinement and analysis of count, mean, and standard deviation of vug areas. Figure 3's distribution plot enhances understanding of vug types, contributing to reservoir characterization. This research not only streamlines vug identification but also advances understanding of subsurface geological formations and vug contribution to hydrocarbon pore space.

**Acknowledgments**

The authors gratefully acknowledge CC Energy Development (CCED), Muscat, Oman, for providing the dataset for this research. The financial support for this project from CCED is also acknowledged. Special appreciation is extended to RealAI for providing the necessary computational resources to facilitate the execution and analysis of the Vugs Identification.